\documentclass[12pt, draftclsnofoot, onecolumn]{IEEEtran}
\usepackage{amsfonts,amssymb}
\usepackage[none]{hyphenat}
\usepackage[ruled,linesnumbered]{algorithm2e}
\usepackage{color}
\usepackage{amsmath}
\usepackage{enumerate}
\usepackage{mathrsfs}
\usepackage{graphicx}
\usepackage{verbatim}
\usepackage{float}
\usepackage{cite}
\newtheorem{Remark}{\it Remark}[section]

\newtheorem{Proposition}{\it Proposition}[section]
\newtheorem{Lemma}{\it Lemma}[section]

\emergencystretch=\maxdimen
\def\BibTeX{\rm{ B\kern-.05em{\sc i\kern-.025em b}\kern-.08em
    T\kern-.1667em\lower.7ex\hbox{E}\kern-.125emX}}
\begin{document}
\title{Jamming Modulation: \\An Active Anti-Jamming Scheme}
\author{Jianhui Ma, Qiang Li, Zilong Liu,~\IEEEmembership{Senior Member,~IEEE}, Linsong Du,  \\ Hongyang Chen,~\IEEEmembership{Senior Member,~IEEE}, and Nirwan Ansari,~\IEEEmembership{Fellow,~IEEE}

\vspace{-5pt}\thanks{ J. Ma and Q. Li are with the  National Key Laboratory of
Science and Technology on Communications, University of Electronic Science and Technology of China (UESTC), Chengdu 611731, China (email: jh.ma16cd@gmail.com; liqiang@uestc.edu.cn). (\emph{Corresponding author: Qiang Li.})

Z. Liu is with the School of Computer Science and Electronics Engineering, University of Essex, Colchester CO4 3SQ, UK (e-mail: zilong.liu@essex.ac.uk).

L. Du is with the School of Information Science and Technology, Southwest Jiaotong University, Chengdu 611756, China (e-mail: linsongdu@outlook.com).

H. Chen is with the Research Center for Graph Computing, Zhejiang Lab, Hangzhou 311121, China (email: dr.h.chen@ieee.org; hongyang@zhejianglab.com). He was supported by key research project of Zhejiang Lab (No. 2022PI0AC01).

N. Ansari is with the Department of Electrical and Computer Engineering, New Jersey Institute of Technology, Newark, NJ 07102 USA (email: nirwan.ansari@njit.edu).
}
}

\markboth{IEEE Transactions on Wireless Communications,~Vol.~XX, No.~XX, XXX~2022}{}
\maketitle

\begin{abstract}
Providing quality communications under adversarial electronic attacks, e.g., broadband jamming attacks, is a challenging task. Unlike state-of-the-art approaches which treat jamming signals as destructive interference, this paper presents  a novel  active anti-jamming (AAJ)  scheme for a jammed channel to enhance the communication quality between a transmitter node (TN) and receiver node (RN), where the TN  actively exploits the jamming signal as a carrier to send messages. Specifically, the TN is equipped with a programmable-gain amplifier, which is capable of  re-modulating the jamming signals for jamming modulation. Considering four typical jamming types,  we derive both the bit error rates (BER) and the corresponding optimal detection thresholds of the AAJ scheme. The asymptotic performances of the AAJ scheme are discussed under the high jamming-to-noise ratio (JNR) and sampling rate cases. Our analysis shows that there exists a BER floor for sufficiently large JNR. Simulation results indicate that the proposed AAJ scheme allows the TN to communicate with the RN reliably even under extremely strong and/or broadband jamming.  Additionally, we investigate the channel capacity of the proposed AAJ scheme, and show that the channel capacity of the AAJ scheme outperforms that of the direct transmission when the JNR is relatively high.
\end{abstract}

\begin{IEEEkeywords}
Anti-jamming communications, jamming modulation (JM), broadband jamming, programmable-gain amplifier (PGA).
\end{IEEEkeywords}
\section{Introduction}
Jamming, also known as an intentional interference \cite{IEEEhowto:1, IEEEhowto:2, IEEEhowto:3}, is one of the most hostile threats in wireless communications. Due to their broadcasting nature, wireless networks are extremely vulnerable to jamming attacks.  A jammer may dramatically disrupt the communications between two or more legitimate users by launching  jamming signals to the target wireless channel \cite{IEEEhowto:4}. As a result, the receiver may fail to decode out any useful messages from the legitimate transmitter.  Since the jamming signal is generally unknown to the receiver, it is difficult to distinguish or suppress them, especially for certain extremely strong and broadband jamming {signals} \cite{IEEEhowto:5}. This motivates us a radical rethinking to re-purpose the notorious jamming signals, instead of passively treating them as destructive interference.

{Anti-jamming and secure communications, {as an important direction in modern telecommunication engineering}, has attracted extensive research attentions over the past decades \cite{IEEEhowto:6,IEEEhowto:7,IEEEhowto:8,IEEEhowto:9,IEEEhowto:10}.} The mainstream anti-jamming techniques include 1) direct sequence spread spectrum (DS-SS) \cite{IEEEhowto:8} by spreading the signal spectrum to suppress the jamming  over a narrowband; 2) frequency hopping (FH) \cite{IEEEhowto:9}  by rapidly switching a carrier over multiple frequency channels by following a pseudo-random hopping pattern known to both the transmitter and receiver  to avoid the jamming attacks; and 3) interference suppression (IS) \cite{IEEEhowto:10} by estimating and  suppressing or abandoning the jammed frequency band. It is noted that these techniques  generally require the legitimate users to have  the jamming types and parameters from the received signal for passive anti-jamming, which is difficult in practice. A major drawback is that the desired signals may be suppressed as well when countering the jamming signals.  In addition, these techniques may not be effective in mitigating extremely strong and/or broadband jamming signals.

Recently, with the emerging  ambient backscatter (AmBC) techniques \cite{IEEEhowto:11}, wireless nodes are able to utilize the existing ambient radio frequency (RF) signals  as a carrier to transmit additional data. As a countermeasure candidate against unknown jamming attacks, the transmitter in an AmBC communication system can modulate the impedance of transmit antenna to change the reflection of an incident signal. To validate its feasibility,  \cite{IEEEhowto:12} investigated the signal detection and bit error rate (BER) performance of  AmBC communication systems. For higher  transmission rates,   \cite{IEEEhowto:13} employed \emph{M}-PSK at the backscatter node to attain high-order modulation. Recently, AmBC has been adopted in \cite{IEEEhowto:14}  to mitigate jamming attacks by developing a reinforcement learning-based algorithm to address the uncertainty of the jamming signals. However, since the backscatter node is assumed to be battery-free and the reflection factor is less than 1, it can neither guarantee a quality communication  nor support a long-distance transmission. To address this problem, we turn to a new solution, i.e., programmable-gain amplifier (PGA), which can amplify the incident signals with programmed gains. According to \cite{IEEEhowto:15}, PGA achieves a broad gain range of $30$ dB over a bandwidth from $2.5$ MHz to $1.17$ GHz. Comparing to AmBC with passive reflecting, PGA is more suitable for anti-jamming communications, especially for broadband jamming cases and high-order modulation.

In this paper, we present a novel active anti-jamming (AAJ) scheme based on jamming modulation (JM). To improve the receiving performance and extend the communication coverage, a PGA is deployed at the transmitter node (TN) to modulate the jamming signals with the varying amplification factors, which are programmable according to the messages to be sent.  Such a scheme  enables us to transmit digital data over jamming signals, instead of counteracting them as in the legacy anti-jamming systems. At the receiver node (RN), the desired messages are retrieved by distinguishing  different energy levels of the received signals. Under this setup, we utilize the signal-to-noise ratio (SNR) and jamming-to-noise ratio (JNR) to characterize the strengths of the desired signal and jamming signal, respectively. The transmission performance is {investigated} to demonstrate the feasibility and reliability of the AAJ scheme.   The main contributions of this paper are summarized as follows:

\begin{itemize}
\item We develop the AAJ scheme to actively  counter the radio jamming attacks by exploiting  the jamming signal as a message carrier. In such a way, AAJ permits legitimate communications even under extremely strong or broadband jamming. Another attractive point of AAJ is that the channel state information (CSI) and the type of  the jamming signal are not required at the receiver, thus making AAJ more practical in real electronic adversarial scenarios.
\item We investigate the BER performance of the AAJ scheme. Both the minimum average BER and the corresponding optimal detection threshold are derived under the energy detection criterion. Moreover, considering the single-tone, multi-tone, narrowband, and broadband jamming types, we derive a general expression of the BER for these four cases.  It is shown that  our analytical BERs match well with the simulated ones. When the JNR under the broadband jamming case is $40$ dB, as an instance shown in Section VII, the BER of the proposed AAJ scheme  is about $7.38\times10^{-5}$. In this case, in sharp contrast, the BERs of the existing schemes approach to $0.5$, leading to a decoding failure, almost surely.

\item We  study the asymptotic behaviors of the BER  under the broadband Gaussian jamming case. In the high JNR case, we prove that there exists a BER floor for sufficiently large JNR; in the high sampling rate case, we derive an approximate expression of the BER.
\item We also investigate the channel capacity of the jammed channel with the AAJ scheme and derive its optimal input distribution in a semi-closed form. Numerical results demonstrate that  the channel capacity of the proposed AAJ scheme exceeds that of the direct transmission (DT) when the JNR is relatively high. More importantly, the system performance improves as the jamming power increases.
\end{itemize}

The remainder of this paper is organized as follows. Section \uppercase\expandafter{\romannumeral2} introduces the system model of the jamming channel. Section \uppercase\expandafter{\romannumeral3} develops an energy detection method and computes the corresponding BER of the proposed AAJ scheme under the random broadband jamming case. Section \uppercase\expandafter{\romannumeral4} analyzes the transmission performances of the AAJ scheme under deterministic jamming cases.  {Section \uppercase\expandafter{\romannumeral5}  shows the asymptotic analysis of the system BER with the AAJ scheme.  Section \uppercase\expandafter{\romannumeral6} investigates the channel capacity  and the corresponding optimal input distribution of the AAJ enabled jamming channel.} Numerical results are presented in Section \uppercase\expandafter{\romannumeral7}. Section \uppercase\expandafter{\romannumeral8} concludes this paper.

\section{System Model}
Let us consider a wireless communications system consisting of a TN, an RN, and a jammer node (JN), as shown in Fig. 1, where the JN transmits  jamming signals to attack both the TN and RN. To enable legitimate communications between the TN and RN, we take advantage of the JM method to introduce a novel AAJ scheme in which the TN is able to send messages over jamming signals.

\subsection{JM Method}
The key idea of {JM is} to utilize the useful messages to modulate the energy of the jamming signals. Specifically, a PGA, whose amplification factor can be {programmable} with digital messages \cite{IEEEhowto:15,IEEEhowto:16,IEEEhowto:17}, is adopted at the TN such that it is able to  communicate with the RN by modulating its own messages over the jamming signals. The JM method can be briefly described in the following three steps:
\begin{itemize}
\item [1)] First, the JN launches a jamming attack, which will be subsequently received by the TN.
\item [2)] Second, the TN   re-modulates and transmits them to the RN by the PGA, which is programmed with the useful digital messages (mapping to different amplification factors)\footnote{The re-modulated signal can be regarded as an energy modulated signal, i.e., different energy levels of the forwarded signal may represent different messages.}.
\item [3)]Finally,  the RN detects the drastic change of the received signals caused by the amplification factors so as to extract the desired messages.
 \end{itemize}

 \begin{figure}[htbp]
 \centering
  \includegraphics[width=0.65\textwidth]{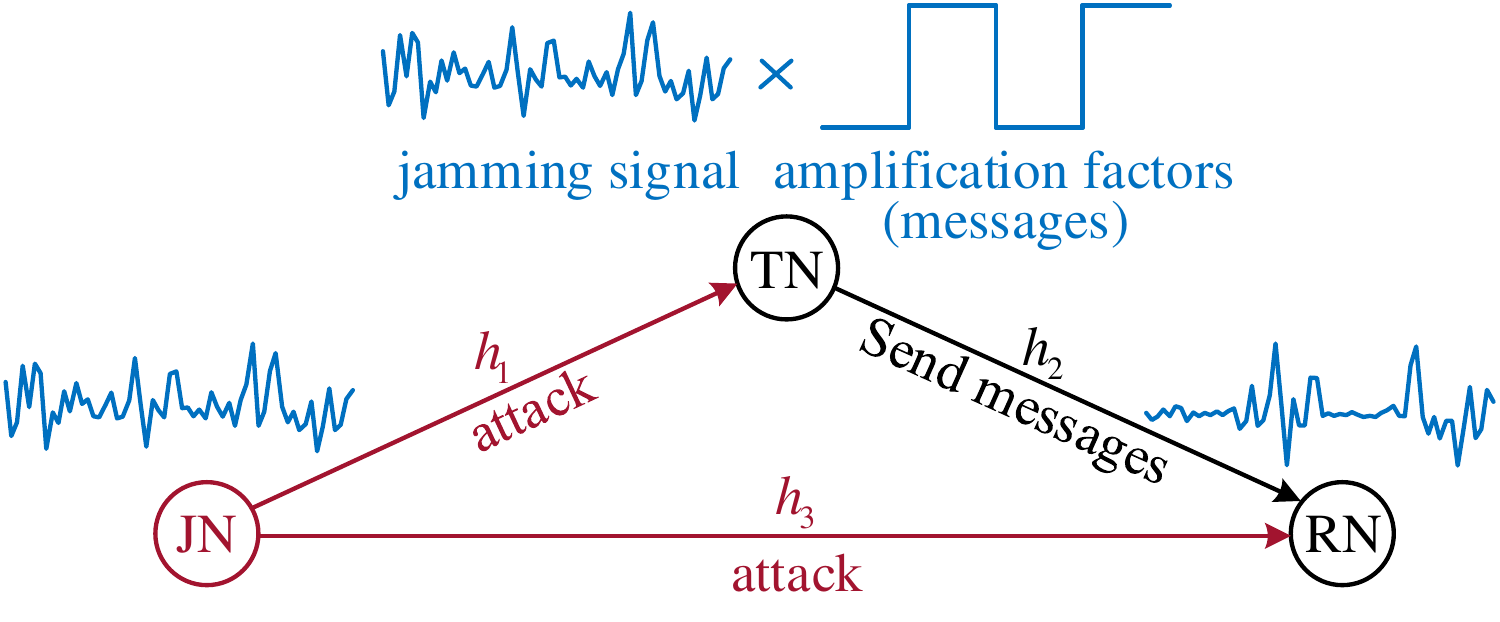}
\caption{Active anti-jamming communications system, with {JN, TN, and RN} denoting the jammer node, transmitter node, and receiver node, respectively, where the {TN} operates in the FD mode.}
\end{figure}
\subsection{Signal Model}
The channel coefficients of the  JN-TN (JT), TN-RN (TR), and  JN-RN (JR) links  are {denoted by} $h_1$, $h_2$, and $h_3$, respectively. In this paper, we consider a frequency-flat and block-fading channel model. All channel coefficients stay constant within a channel coherent time but may vary independently in different coherent intervals. In real electronic adversarial cases,  the CSI of the jamming channel, i.e., JT and JR links,  and the type of jamming signal are generally unavailable. Therefore, the channel coefficients $h_1$ and $h_2$ as well as the jamming signal {may} be unknown to the TN and RN. Without any loss of generality, the TN operates in the full-duplex (FD) mode, which enables it to receive and transmit signals at the same time slot. Next, the signal transmitting and receiving processes  of  the proposed AAJ scheme are described as follows.

\subsubsection{Signal Transmitting} The JN broadcasts the jamming signals to both the TN and RN\footnote{{In this paper, we focus on the case {of continuous jamming attack, i.e., the jammer keeps transmitting without a stop.} The jamming signal can be detected by the spectrum sensing technique  \cite{IEEEhowto:18,IEEEhowto:19}.  If  jamming is not detected, the TN may switch to the conventional RF chain to transmit signals, i.e., generating signal carrier by itself.}}.
Therefore, the received signal $Y_{\rm {T}}(t)$  at the TN is given by
\begin{align}
Y_{\rm{ T}}(t)&=h_1 X_{\rm J}(t).
\end{align}
Here, $X_{\rm J}(t)$ denotes the unknown jamming signal. For the convenience of calculation, the noise of the PGA and the self-interference introduced by the FD mode at the TN are both incorporated into the noise at the RN\footnote{{Unlike conventional communications, the  digital-to-analog converter (DAC), the oscillator, and the  frequency mixer, etc., are not required at the TN when the AAJ scheme is adopted, and thus the  noise at the TN is relatively low. In addition, according to the current self-interference cancellation techniques, the residual self-interference is also very low as compared with the strong jamming signal \cite{IEEEhowto:20},\cite{IEEEhowto:21}. Therefore, both of them can be regarded as background noises when the RN decodes the messages.}}.

Under this case, the TN communicates with the RN by adopting the JM method, i.e., the messages to be sent by the TN are mapped to different amplification factors, carried by the jamming signal, and  forwarded to the RN. Therefore, the transmitted signal  $X_{\rm T}(t)$ at the TN is given by
\begin{align}
X_{\rm T}(t)&=A(t)Y_{\rm T}(t)\notag\\
&=A(t)h_1 X_{\rm J}(t),
\end{align}
where $A(t)\in\{a_k\}, k=1,2,...,K$, is the amplification factor corresponding to the messages to be sent by the TN  with  the distribution $p(A(t)=a_k)=p_k$, $\sum^{K}_{k=1}p_{k}=1$.

\subsubsection{Signal Receiving} {As shown in Fig. 1, the RN receives the superposition of the jamming signal  from the JR link and the re-modulated signal from the TR link}. Therefore, the received signal $Y_{\rm R}(t)$ at the RN is given by
\begin{align}
Y_{{\rm R}}(t)&=h_2X_{\rm T}(t)+h_3X_{\rm J}(t-\tau)+Z_{\rm R}(t)\notag\\
&=h_1h_2A(t)X_{\rm J}(t)+h_3X_{\rm J}(t-\tau)+Z_{\rm R}(t),
\end{align}
where $\tau$ is the delay between the JR and JT-TR links, and $Z_{\rm R}(t)$ is the independent and identical distributed (i.i.d.) circularly symmetric complex Gaussian (CSCG)  noise, i.e.,  $Z_{\rm R}(t)\sim\mathcal{CN}(0,\sigma^2_{\rm R})$.

\begin{Remark}
It is observed that the received signal at the RN  includes not only the additive interference and noise $h_3X_{\rm J}(t-\tau)+Z_{\rm R}(t)$ but also the multiplicative interference $h_1h_2A(t)X_{\rm J}(t)$, thus making the decoding a complex task.
\end{Remark}
\begin{Remark}
Note that from (3), the desired message, i.e., $A(t)$,  is  only carried by the jamming signal received from the TR link. In other words, the jamming signal from the JR link, i.e., $h_3X_{\rm J}(t-\tau)$, is regarded as interference when the RN decodes the desired messages.
\end{Remark}

\section{Proposed Signal Detection}
In this section,  an energy detector \cite{IEEEhowto:22} is introduced  for  the proposed AAJ scheme under the jamming attacks. Unlike conventional communications, the CSI and modulation type of the hostile jammer in  anti-jamming communications are generally unknown to the receiver, and thus coherent demodulations are inapplicable in this case. However, some statistics of the jamming signal are generally fixed due to the hardware limits of the jammer. For instance, the average transmission power of the jammer is fixed\footnote{{It is worth noting that the proposed AAJ scheme is not limited to the fixed jamming power and signal type cases. Although the transmission performances may degrade,  it can still work  even under jamming cases with varying power levels due to the fact that the signal energy from the JT-TR link can be dramatically changed by the amplification factor $a_k$. The transmission performances of the AAJ scheme under the other jamming cases are thus worthy of a further study.}}. This inherent property has led to the proposed energy-based detection scheme, which only relies on the statistics of the received signal to recover the desired messages.

During the duration of one transmitted symbol, the received signal at the RN is sampled for $N$ times, which is denoted {by} ${\boldsymbol{y}}_{_{\rm R}}=[y_{_{\rm R}}[1], y_{_{\rm R}}[2],...,y_{_{\rm R}}[N]]^T$.
From (3), when the TN sends the symbol $a_k$, the $n$-th sample can be expressed as
\begin{align}
y_{_{\rm R}}[n]=h[n]a_k+z[n],
\end{align}
where $h[n]=h_1h_2X_{\rm J}[n]$ and $z[n]=h_3X_{\rm J}[n-n_{\tau}]+Z_{\rm R}[n]$.

According to (4), the signal-to-interference and noise ratio (SINR) at the RN with the AAJ scheme can be given by
\begin{align}
{\rm SINR}&=\frac{ \mathbb{E}\left[ \left\vert h_1h_2X_{\rm J}[n]a_k\right\vert^2\right]}{\mathbb{E} \left[\left\vert h_3X_{\rm J}[n-n_{\tau}]+Z_{\rm R}[n]\right\vert^2\right]}\notag\\
&=\frac{|h_1|^2|h_2|^2|a_k|^2P_{\rm J}}{|h_3|^2P_{\rm J}+\sigma^2_{\rm R}},
\end{align}
where $P_{\rm J}$ is the average power of the jamming signal.
\begin{Remark}
One can see that  unlike conventional anti-jamming communications, both the numerator and the denominator of the SINR include the term of $P_{\rm J}$ in the AAJ communications system due to the fact that we exploit the jamming signal to transmit messages\footnote{{As a future work, one may consider an adaptive scheme in terms of the jamming strength, where $A(t)$ depends on the strength of $X_{\rm J}(t)$. When the jamming is strong, we can choose a smaller $A(t)$ to maintain the energy level of the re-modulated jamming signal; when the jamming is weak, we should choose a larger $A(t)$ or switch to the conventional anti-jamming scheme.}}.
\end{Remark}
\begin{Remark}
Since the jamming signal $X_{\rm J} [n]$ is unknown to the RN, it may not directly extract the desired messages from the received signal. However,  considering the limits of hardware, the JN is supposed to launch jamming attacks with the fixed  transmission power. Consequently, the average power of the jamming signal is mostly fixed, e.g., the single-tone jamming signal with a constant envelope.
\end{Remark}

\subsection{Energy Detection}
\subsubsection{Energy Detector}
An energy detector measures the average energy of the received signal over one transmitted symbol period and compares it with a predefined threshold to distinguish the different symbols. The average energy of $N$ receiver samples corresponding to one transmit symbol, i.e., the average energy of ${\boldsymbol{y}}_{_{\rm R}}$, is  given by
\begin{align}
{\mathcal{Q}}\triangleq\frac{1}{N} \sum^N_{n=1} |y_{_{\rm R}}[n]|^2.
\end{align}

By substituting (4) into (6), {we have}
\begin{align}
\frac{1}{N}\sum^{N}_{n=1}\big|y_{_{\rm R}}[n]\big|^2=&\frac{1}{N}\sum^{N}_{n=1}\big|h_1h_2a_k X_{\rm J}[n]+h_3X_{\rm J}[n-n_{ \tau}]+Z_{\rm R} [n]\big|^2\notag\\
=&\frac{1}{N}\sum^{N}_{n=1}\big|h_1h_2a_k X_{\rm J}[n]+h_3X_{\rm J}[n-n_{ \tau}]\big|^2+\frac{1}{N}\sum^{N}_{n=1}\big|Z_{\rm R}[n]\big|^2,
\end{align}
where (7) is obtained by leveraging the independence between the jamming signal and noise.

 {For sufficiently large $N$, we have
 \begin{align}
 \frac{1}{N}\sum^{N}_{n=1}\big|h_1h_2a_k X_{\rm J}[n]+h_3X_{\rm J}[n-n_{ \tau}]\big|^2\longrightarrow\big|h_1h_2a_k+h_3\big|^2\frac{1}{N}\sum^{N}_{n=1}\big|X_{\rm J}[n]\big|^2.
  \end{align}
Note that we focus on the strong jamming case in this paper. {By ignoring the noise term}, (7) can be approximated by
\begin{align}
{\mathcal{Q}}&=\frac{1}{N} \sum^N_{n=1} |y_{_{\rm R}}[n]|^2\notag\\
&\approx\big|h_1h_2a_k+h_3\big|^2\frac{1}{N}\sum^{N}_{n=1}\big|X_{\rm J}[n]\big|^2.
\end{align}
Therefore, the average energy of the received signal is determined by the amplification factor $a_k$, meaning that the RN can extract the desired message from the received signal by  distinguishing its different energy levels.}
{\begin{Remark}
 For instance, if we adopt the binary modulation, i.e., on-off keying (OOK), at the TN, we have
 \begin{align}
{\mathcal{Q}}\approx\left\{
             \begin{array}{lr}
            \big|h_3\big|^2\frac{1}{N}\sum^{N}_{n=1}\big|X_{\rm J}[n]\big|^2, & a_1=0 \ \text{for sending `0'},\\
          \big|h_1h_2a+h_3\big|^2\frac{1}{N}\sum^{N}_{n=1}\big|X_{\rm J}[n]\big|^2,  &  a_2=a \ \text{for sending `1'}.
             \end{array}
             \right.
 \end{align}
 Obviously,  the energy levels  of the received signal are distinguishable at the RN  as long as $a$ is sufficiently large.
\end{Remark}}

\subsubsection{The Estimation of Detection Threshold}The threshold value $\mathcal{T}_{\rm th}$ for detecting a symbol is derived by a preamble promised beforehand, e.g., `101010', in the case of binary input. When the RN receives the whole signal during one communication block, it first uses the preamble sequence to determine the average energy of `0' and `1', and then sets the threshold. The average energy of the two preamble symbols are given by
\begin{align}
\hat{{\mathcal{Q}}}_p=\left\{
             \begin{array}{lr}
            \left(P_2+P_4+P_6\right)/3, & \text{preamble symbol is `0'},\\
            \left(P_1+P_3+P_5\right)/3,  & \text{preamble symbol is `1'},
             \end{array}
             \right.
\end{align}
where $P_m$ denotes the average energy of the $m$-th symbol in the preamble, i.e.,
\begin{align}
P_m=\frac{1}{N}\sum^{N}_{n=1}\big|y_{_{\rm R}}[(m-1)N+n]\big|^2, 1<m<M.
\end{align}
In practice, we can choose a  longer preamble, i.e., $M$ is relatively large, to get a more precise threshold.

Therefore, the detection threshold between `0' and `1'  can be formulated as
\begin{align}
\hat{\mathcal{T}}_{\rm th}&=T_{\rm ED}\left(\hat{{\mathcal{Q}}}_p(a_k=a_1),\hat{{\mathcal{Q}}}_p(a_k=a_2)\right),
\end{align}
where $T_{\rm ED}$ is the mapping from $\hat{\mathcal{Q}}_p$ to $\hat{\mathcal{T}}_{\rm th}$, which is given in {\emph{Remark 3.5}} of the next subsection. {$\hat{{\mathcal{Q}}}_p(a_k=a_1)$ and $\hat{{\mathcal{Q}}}_p(a_k=a_2)$ correspond to the estimated average energy levels of `0' and `1', respectively, which are given in (11).}

\subsubsection{Message Decoding} $\hat{\mathcal{T}}_{\rm th}$
 divides the estimation space into non-intersection intervals, i.e., $\mathcal{I}_1$ and $\mathcal{I}_2$, corresponding to different amplification factors  $\{a_k\}_{k=1}^2$, i.e.,
\begin{align}
\mathcal{I}_k=\left\{
             \begin{array}{lr}
              {\mathcal{Q}}\in(0,\hat{\mathcal{T}}_{\rm th}),  &  a_k=a_1 \ \text{for sending `0'}, \\
            {\mathcal{Q}}\in(\hat{\mathcal{T}}_{\rm th},+\infty), & a_k=a_2\ \text{for sending `1'}.
             \end{array}
             \right.
\end{align}

Hence, the decoder at the RN  produces
\begin{align}
\hat{k}:\big\{k\in\{1,2\}: {\mathcal{Q}}\in\mathcal{I}_k\big\}.
\end{align}

Next, let us  summarize the energy detection method in the following three steps:
\begin{enumerate}[\quad\bf{Step}\ 1:]
\item {Compute} the average energy of received signal $\mathcal{Q}$ by (6);
\item {Estimate} the detection threshold $\hat{\mathcal{T}}_{\rm th}$ by (11) and (13);
\item {Decode} the desired messages by the energy detection criterion. If $\mathcal{Q}\leq\hat{\mathcal{T}_{\rm th}}$, then $a_k=a_1$; else $a_k=a_2$.
\end{enumerate}

\begin{Remark}
  It is worth noting that the energy detection, as {a non-coherent} demodulation, only uses the average energy of the received signal, i.e., $\mathcal{Q}=\frac{1}{N} \sum^N_{n=1} |y_{_{\rm R}}[n]|^2$,  to decode the desired messages. Consequently, the CSI of the anti-jamming channel and even the type of jamming signal {may not be necessary}  for message recovering, which is the very unique feature of the AAJ scheme in practice.
\end{Remark}
\subsection{Bit Error Rate}
To estimate the performance of the proposed AAJ scheme, we next compute its average BER, which is defined as
\begin{align}
P_e=\sum^K_{k=1} P_e(a_k),
\end{align}
where $P_e(a_k)=p_k\left(1-\int_{\mathcal{I}_k}f(q){\rm d}q\right)$ is the error probability when the TN transmits symbol $a_k$, and $f(q)$ is the probability density function (PDF) of ${\mathcal{Q}}$, i.e., the average energy of $N$ receiver samples.

Since the transmit power budget of the jammer transmitter is limited, we have
\begin{align}
\mathbb{E}\left[|X_{\rm J}[n]|^2\right]&=\lim\limits_{N\to \infty} \frac{1}{N}\sum\limits^{N}_{n=1}|X_{\rm J} [n]|^2\leq P_{\rm J}.
\end{align}

According to the maximum entropy principle \cite{IEEEhowto:23}, the Gaussian distribution maximizes the entropy over all distributions under the same variance. Therefore,  the uncertainty of the jamming signal achieves the maximum value when it subjects to the zero-mean complex Gaussian distribution, i.e., $X_{\rm J}[n]\sim \mathcal{CN}(0,P_{\rm J})$. In other words, the best strategy of the JN is to launch a Gaussian jamming signal.

In such a case, the RN received signal under $A[n]=a_k$,  is also a zero-mean complex Gaussian random process, i.e.,
\begin{align}
y_{_{\rm R}}[n]\sim\mathcal{CN}\left(0,\delta^{2}_{k}\right),
\end{align}
where the delay at the TN is omitted, and the variance  of $y_{_{\rm R}}[n]$ is computed as
\begin{align}
\delta^{2}_{k}&={\rm Var}\left[h[n]a_k+z[n]\right]\notag\\
&=\left\vert h_1h_2a_k+h_3\right\vert^2 P_{\rm J}+\sigma^2_{\rm R}.
\end{align}

It can be observed from (6) that ${\mathcal{Q}}$ is a sum of squares of $N$ independent CSCG variables with identical mean and variance. By normalizing the variance of $y_{_{\rm R}}[n]$, we have the following lemma.

\begin{Lemma}
The average energy of the received signal during one transmitted symbol ${\mathcal{Q}}$ satisfies
\begin{align}
\frac{2N}{\delta^2_k}{\mathcal{Q}}\sim\chi^2(2N),
\end{align}
where $\chi^2(2N)$ is the chi-squared distribution with $2N$ degrees of freedom.
\end{Lemma}

\emph{Proof:} Please see Appendix A.\hfill\rule{2mm}{2mm}

According to {\emph{Lemma 3.1}}, the distribution of $(2{\mathcal{Q}}N) / \delta^2_k$ is derived, upon which we can further compute the PDF of ${\mathcal{Q}}$.

\begin{Lemma}
The PDF of  ${\mathcal{Q}}$ is  given {by}
\begin{align}
f_{\mathcal{Q}}(q)=\left\{
             \begin{array}{lr}
              \frac{N}{\Gamma(N)\left(\delta^2_k\right)^N} \left(N q\right)^{N-1} e^{-\frac{N}{\delta^2_k}q},  & q>0,\\
           0, & \text{otherwise},
             \end{array}
             \right.
\end{align}
where $\Gamma(N)=\int_0^{+\infty} x^{N-1}e^{-x}{\rm d}x$ is the Gamma function.
\end{Lemma}

\emph{Proof:} Please see Appendix B.\hfill\rule{2mm}{2mm}

By substituting the above PDF of $\mathcal{Q}$ into (16), we can compute the optimal detection threshold for the energy detector, which is given {by}
\begin{align}
\mathcal{T}^*_{\rm th}=\arg\min\limits_{\mathcal{T}_{\rm th}} P_e.
\end{align}

Considering the binary input at the TN and  $\delta^2_1<\delta^2_2$, the minimum $P_e$ and the corresponding optimal detection threshold $\mathcal{T}_{\rm th}^*$  of the proposed AAJ scheme under the energy detection is {summarized} in the following proposition.

\begin{Proposition}
The minimum average BER of  the proposed AAJ scheme under the energy detection can be expressed as
\begin{align}
P_e=&p_1 \left(1-\int^{\mathcal{T}^*_{\rm th}}_0  \frac{N}{\Gamma(N)\left(\delta^2_1\right)^N} \left(N q\right)^{N-1} e^{-\frac{N}{\delta^2_1}q} \right){\rm d}q\notag\\
&+p_2 \left(1-\int_{\mathcal{T}^*_{\rm th}}^{+\infty}  \frac{N}{\Gamma(N)\left(\delta^2_2\right)^N} \left(N q\right)^{N-1} e^{-\frac{N}{\delta^2_2}q} \right){\rm d}q,
\end{align}
where the optimal detection threshold $\mathcal{T}^*_{\rm th}$ is determined by
\begin{align}
\mathcal{T}^*_{\rm th}=\frac{1}{N}\frac{\delta^2_1 \delta^2_2}{\delta^2_2-\delta^2_1}\ln \left(\frac{p_1}{p_2}\left(\frac{\delta^2_2}{\delta^2_1}\right)^N\right).
\end{align}
\end{Proposition}

\emph{Proof:} Please see Appendix C. \hfill\rule{2mm}{2mm}

\begin{Remark}
Since $y_{_{\rm R}}[n]$ is a zero mean variable, and then we have $\delta_k^2=\mathbb{E}\left[|y_{_{\rm R}}[n]|^2\right]\to \hat{\mathcal{Q}}_p$ when $N$ is large enough.  From (13) and (24),  the mapping $T_{\rm ED}$ is determined by $T_{\rm ED}(x,y)=$ $\frac{1}{N}\frac{x y}{y-x}\ln \left(\frac{p_1}{p_2}\left(\frac{y}{x}\right)^N\right)$. Consequently, when $p_1=p_2=0.5$, the optimal detection threshold  in practice becomes
\begin{align}
\mathcal{T}^*_{\rm th}=\frac{\hat{\mathcal{Q}}_p(a_k=a_1) \hat{\mathcal{Q}}_p(a_k=a_2) }{\hat{\mathcal{Q}}_p(a_k=a_2) -\hat{\mathcal{Q}}_p(a_k=a_1) }\ln \frac{\hat{\mathcal{Q}}_p(a_k=a_2) }{\hat{\mathcal{Q}}_p(a_k=a_1) }.
\end{align}
It is obvious that the optimal detection threshold only relates to the average energy of the received signal samples, which can be estimated by $\hat{\mathcal{Q}}_p$.
\end{Remark}

\section{Deterministic jamming case}
In the {above discussions, we have analyzed} the error performance of the AAJ scheme under the Gaussian jamming signal, which can be regarded as a type of random broadband jamming\footnote{The narrowband Gaussian jamming signal can be generated by a white Gaussian signal passing through a bandpass filter. However, the filtering operation brings correlation among the sampled symbols, which does not fit the signal model of this paper. Therefore, we will discuss this case in our next paper.}. In this subsection, we investigate the BERs of the AAJ scheme under the deterministic jamming, which includes single-tone, multi-tone, narrowband, and broadband jamming signals. Moreover, the general expression of the system BER under the above five jamming cases is derived.
\subsection{Single-Tone Jamming} The JN launches a single-tone jamming signal, i.e., a sinusoid signal with a certain frequency, which can be expressed as
\begin{align}
X_{\rm J}(t)=a_ j\cos(2\pi f_j t+\phi_j),
\end{align}
where $a_ j$, $f_j$, and $\phi_j$ are the amplitude, the carrier frequency, and  the initial phase of the jamming signal, respectively.

From (4), the received signal after the sampling is given by
\begin{align}
y_{_{\rm R}}[n]=&h_1h_2a_k a_ j\cos[2\pi f_j n+\phi_j]\notag\\
&+h_3a_ j\cos[2\pi f_j (n-n_{\tau})+\phi_j] +Z_{\rm R}[n].
\end{align}
\subsection{Multi-Tone/Narrowband/Broadband Jamming}  The deterministic multi-tone, narrowband, and broadband jamming signals can be regarded as  a sum of sinusoid signals. The difference among these three jamming types are summarized as follows:
\begin{enumerate}
\item Multi-tone jamming: the frequencies of the multi-tone jamming signal  can be arbitrarily distributed in the whole band.
\item Narrowband jamming: the frequencies of the narrowband jamming signal occupy a narrow frequency range.
\item Broadband jamming: the frequencies of the broadband jamming signal occupy a broadband frequency range.
\end{enumerate}

Generally, the bandwidth of the narrowband jamming signal is less than 1$\%$ of that of the source signal while  the bandwidth of the broadband jamming signal is larger than 10$\%$ of that of the source signal. All these three jamming signals can be expressed as
\begin{align}
X_{\rm J}(t)=\sum^{J}_{j=1}a_ j\cos(2\pi f_j t+\phi_j).
\end{align}

Consequently, the received signal after the sampling is given by
\vspace{-10pt}
\begin{align}
y_{_{\rm R}}[n]=&h_1h_2a_k \sum^{J}_{j=1}a_ j\cos[2\pi f_j n+\phi_j]\notag\\
&+h_3\sum^{J}_{j=1}a_ j\cos[2\pi f_j (n-n_{\tau})+\phi_j] +Z_{\rm R}[n].
\end{align}

Further, under the deterministic single-tone, multi-tone, narrowband, and broadband jamming cases, the general expression of the average energy for $N$ received samples is given by
\begin{align}
\mathcal{Q}=&\underbrace{\frac{1}{N} \sum^N_{n=1}\left\vert h_1h_2a_k \sum^{J}_{j=1}a_ j\cos[2\pi f_j n+\phi_j]+h_3\sum^{J}_{j=1}a_ j\cos[2\pi f_j (n-n_{\tau})+\phi_j]\right\vert^2}_{\text  { \normalsize $\mathcal{Q}_{{\rm D},k}$: Deterministic part}}\notag\\
&+\underbrace{\frac{1}{N} \sum^N_{n=1}\left\vert Z_{\rm R}[n]\right\vert^2.}_{\text {\normalsize $\mathcal{Q}_{{\rm R},k}$: Random part}}
\end{align}
\subsection{General Expression of BER}
It is worth noting that when $N$ is large enough, the average energy of the deterministic signal part  in (30), denoted by $\mathcal{Q}_{{\rm D},k}$, is close to a constant. The uncertainty of $\mathcal{Q}$ mainly comes from the noise part, denoted by $\mathcal{Q}_{{\rm R},k}$. According to \emph{Lemma 3.2}, $\mathcal{Q}_{{\rm R},k}$ is a sum of squares of $N$ complex Gaussian signals, with the following PDF
\begin{align}
f_{\mathcal{Q}_{{\rm R},k}}(q)=\left\{
             \begin{array}{lr}
              \frac{N}{\Gamma(N)\left(\sigma^2_{\rm R} \right)^N} \left(N q\right)^{N-1} e^{-\frac{N}{\sigma^2_{\rm R}}q},  & q>0,\\
           0, & \text{otherwise}.
             \end{array}
             \right.
\end{align}

Since $\mathcal{Q}=\mathcal{Q}_{{\rm D},k}+\mathcal{Q}_{{\rm R},k}$, we have
\begin{align}
F_{\mathcal{Q}}(q)&={\rm Pr}\{\mathcal{Q}\leq q\}\notag\\
&={\rm Pr}\left\{\mathcal{Q}_{{\rm R},k}\leq q-\mathcal{Q}_{{\rm D},k}\right\}\notag\\
&=F_{\mathcal{Q}_{2,k}}\left(q-\mathcal{Q}_{{\rm D},k}\right).
\end{align}

Therefore, the PDF of $\mathcal{Q}$ can be computed by
\begin{align}
f_{\mathcal{Q}}(q)&=\frac{{\rm d}F_{\mathcal{Q}}(q)}{{\rm d} q}=f_{\mathcal{Q}_{{\rm R},k}}(q-\mathcal{Q}_{{\rm D},k}).
\end{align}
\indent By substituting $f_{\mathcal{Q}_{{\rm R},k}}(q)$ into (33), we have
\begin{align}
f_{\mathcal{Q}}(q)=\left\{
             \begin{array}{lr}
              \frac{N}{\Gamma(N)\left(\sigma^2_{\rm R} \right)^N} \left(N (q-\mathcal{Q}_{{\rm D},k})\right)^{N-1} e^{-\frac{N}{\sigma^2_{\rm R}}(q-\mathcal{Q}_{{\rm D},k})},  & q>0,\\
           0, & \text{otherwise}.
             \end{array}
             \right.
\end{align}

The minimum $P_e$ and the corresponding optimal $\mathcal{T}_{\rm th}^*$  of  the proposed AAJ scheme under the deterministic jamming case is summarized in the following proposition.

\begin{Proposition}
The minimum BER of  the proposed AAJ scheme under the deterministic jamming case can be expressed as
\begin{align}
P_e=&p_1 \left(1-\int^{\mathcal{T}_{\rm th}^*}_0  \frac{N}{\Gamma(N)\left(\sigma^2_{\rm R} \right)^N} \left(N (q-\mathcal{Q}_{{\rm D},1})\right)^{N-1} e^{-\frac{N}{\sigma^2_{\rm R}}(q-\mathcal{Q}_{{\rm D},1})} \right){\rm d}q\notag\\
&+p_2 \left(1-\int_{\mathcal{T}_{\rm th}^*}^{+\infty}  \frac{N}{\Gamma(N)\left(\sigma^2_{\rm R} \right)^N} \left(N(q-\mathcal{Q}_{{\rm D},2})\right)^{N-1} e^{-\frac{N}{\sigma^2_{\rm R}}(q-\mathcal{Q}_{{\rm D},2})} \right){\rm d}q,
\end{align}
where the optimal detection threshold $\mathcal{T}^*_{\rm th}$ is determined by
\begin{align}
\mathcal{T}^*_{\rm th}=\frac{\mathcal{Q}_{{\rm D},1}-\xi\mathcal{Q}_{{\rm D},2}}{1-\xi}.
\end{align}
Here, $\mathcal{Q}_{{\rm D},1}<\mathcal{Q}_{{\rm D},2}$ and $\xi=\frac{p_2}{p_1}e^{\frac{N-1}{N}\left(\frac{\mathcal{Q}_{{\rm D},2}-\mathcal{Q}_{{\rm D},1}}{\sigma_{\rm R}^2}\right)}$.
\end{Proposition}

\emph{Proof:} The proof is similar to that of \emph{Proposition 3.1}, and thus it is omitted for brevity. \hfill\rule{2mm}{2mm}

Although $\mathcal{Q}_{{\rm D},1}$ and $\mathcal{Q}_{{\rm D},2}$ in (36) cannot be derived directly, $\mathcal{Q}_{{\rm D},2}-\mathcal{Q}_{{\rm D},1}$ can be estimated by $\hat{\mathcal{Q}}_p(a_k=a_2)$ and $\hat{\mathcal{Q}}_p(a_k=a_1)$ such that $\mathcal{T}^*_{\rm th}$ can also be derived.

Notice that the BERs of the  single-tone, multi-tone, narrowband, and broadband jamming (random and deterministic types) cases have a similar structure. Thus, we develop a general expression of the BER for the considered five jamming cases, i.e.,
\begin{align}
P_e\!=\!\sum_{k=1}^K p_k \!\left(\!1\!-\!\int_{\mathcal{I}_k} \frac{N}{\Gamma(N)A^N} \left(N (q-B)\right)^{N-1} e^{\!-\!\frac{N}{A}(q-B)} \right){\rm d}q.
\end{align}

\begin{enumerate}[\quad 1)]
\item When $A=\delta^2_k$ and $B=0$, $P_e$ is the BER of the random broadband jamming case.
\item When $A=\sigma^2_{\rm R}$ and $B=\mathcal{Q}_{{\rm D},k}$, $P_e$ is the BER of the deterministic single-tone, multi-tone, narrow-band, or broadband jamming case.
\end{enumerate}
\begin{Remark}
Since the deterministic broadband, narrowband, and multi-tone jamming signals can be regarded as a sum of different single-tone jamming signals, the BER of the single-tone jamming case is generally lower than the other three cases. The reason is that the different single-tone jamming signals may cancel each other out, resulting in a decrease of signal strength. Therefore, under the same transmission power and channel conditions, the BER performances of the above four cases satisfy: $P_e^{\rm ST}\leq P_e^{\rm MT}\leq P_e^{\rm RB}\leq P_e^{\rm BB,D}\leq P_{e}^{\rm BB,R}$, where $P_e^{\rm ST}$, $P_e^{\rm MT}$, $P_e^{\rm RB}$, and $P_e^{\rm BB, D}$  are the BERs of the deterministic single-tone, multi-tone, narrowband, and broadband jamming cases, respectively, and $P_{e}^{\rm BB, R}$ is the BER of the random broadband jamming case.
\end{Remark}

{\section{Asymptotic Analysis}
To uncover more insights on the error performance  of the AAJ  enabled anti-jamming communications, we investigate the asymptotic behaviors of the BER, where the power of the jamming signal and the number of samples go to infinity. Here, we define the JNR as $P_{\rm J}/\sigma^2_{\rm R}$ to characterize the effect of the power of the jamming signal.
\subsection{The High JNR Case}
Recalling (5), we can {rewrite} the SINR at the RN as {follows:}
\begin{align}
{\rm SINR}=\frac{|h_1|^2|h_2|^2|a_k|^2}{|h_3|^2+\sigma^2_{\rm R}/P_{\rm J}}.
\end{align}
It can be observed from (38) that the SINR monotonically increases with the JNR, meaning that the more power the JN uses, the higher SINR the RN can get, i.e.,  the better BER the system can obtain.

\begin{Proposition}
If $a_k$ is fixed and $P_{\rm J}/\sigma_{\rm R}^2\rightarrow \infty$, SINR increases to $\left(|h_1|^2|h_2|^2|a_k|^2\right)/|h_3|^2$ at the following rate
\begin{align}
{\rm SINR}\approx O\left(1\right).
\end{align}
\end{Proposition}

\emph{Proof:} When $P_{\rm J}/\sigma_{\rm R}^2\rightarrow \infty$, $\sigma^2_{\rm R}/P_{\rm J}\rightarrow0$. From (38), we have
\begin{align}
\left.{\rm SINR}\right|_{P_{\rm J}/\sigma_{\rm R}^2\rightarrow\infty}\longrightarrow \frac{|h_1|^2|h_2|^2|a_k|^2}{|h_3|^2},
\end{align}
which is $O\left(1\right)$. \hfill\rule{2mm}{2mm}
\begin{Remark}
Based on the above analysis, we can conclude that the SINR  increases and approaches  the upper bound with the increase of the JNR. In this case, the BER  decreases and approaches  the lower bound with the increase of the JNR. In other words, there will be a BER floor with the increasing JNR. In addition, the lower bound of BER is determined by  the channel conditions and the transmission power of the TN.
\end{Remark}
\begin{Remark}
As we know, SINR is determined by the ratio of  the strength of desired signal  to the variances of  jamming signal and noise.  The SINR expression (38) holds when  $X_{\rm J}[n]$ is a random signal.  If $X_{\rm J}[n]$ is the deterministic single-tone, multi-tone,  narrowband, or broadband jamming signal, the variance of the jamming signal ${\rm Var} \left[X_{\rm J}[n]\right]=0$. Therefore, the SINRs of the  above four deterministic jamming cases are given by
\begin{align}
{\rm SINR}=\frac{|h_1|^2|h_2|^2|a_k|^2P_{\rm J}}{\sigma^2_{\rm R}}.
\end{align}
{Unlike} the random broadband jamming case, the power of the jamming signal $P_{\rm J}$ in (41) is not in the denominator of the SINR when the JN launches the deterministic jamming signals. In other words, the SINR monotonically increases with $P_{\rm J}$, i.e., the BER monotonically decreases with $P_{\rm J}$ under the deterministic jamming  cases.
\end{Remark}
\subsection{The High Sampling Rate Case}

We now look into the asymptotic regime, where the number of samples $N$ is large during one symbol, i.e., the high sampling rate. It is observed from (18) that $|y_{_{\rm R}}[n]|^2$ is a central chi-square random variable with two degrees of freedom, which follows $\left|\sqrt{2}y_{_{\rm R}}[n]/\delta_k\right|^2\sim \mathcal{X}^2(2)$. From the statistical property of the chi-square distribution, the mean and variance of $\left|y_{_{\rm R}}[n]\right|^2$ are $\delta_k^2$ and $\delta_k^4$, respectively. Consequently, $\mathcal{Q}=\frac{1}{N}\sum_{n=1}^{N}|y_{_{\rm R}}[n]|^2$ is a sum of $N$ independent identical distributed central chi-square variables with two degrees of freedom, and  we can derive the following proposition:

\begin{Proposition}
For sufficiently large $N$,  the average energy of the sampled signal $\mathcal{Q}$ approximately follows the Gaussian distribution, i.e.,
\begin{align}
\mathcal{Q}\sim\mathcal{N}\left(\delta^2_k,\delta^4_k/N\right).
\end{align}
\end{Proposition}

{\emph{Proof:} According to the central limit theorem \cite{IEEEhowto:24}, for sufficiently large $N$, we have
\begin{align}
\frac{\frac{1}{N}\sum_{n=1}^N |y_{_{\rm R}}[n]|^2-\mu}{\sigma/\sqrt{N}}&\sim \mathcal{N}(0,1)\notag\\
\longrightarrow \mathcal{Q}=\frac{1}{N}\sum_{n=1}^N |y_{_{\rm R}}[n]|^2&\sim\mathcal{N}(\mu,\sigma^2/N),
\end{align}
where $\mu=\sigma=\delta^2_k$. Hence, \emph{Proposition 5.2} is proved.\hfill\rule{2mm}{2mm}

Therefore, the BER in (23) can be rewritten as
\begin{align}
P_e=&p_1P_e(a_1)+p_2P_e(a_2)\notag\\
\approx& p_1\int_{\mathcal{T}_{\rm th}^*}^{+\infty}\frac{1}{\sqrt{2\pi\delta_1^4/N}}e^{-\frac{\left(q-\delta^2_1\right)^2}{2\delta^4_1/N}}{\rm d} q+p_2\int_{0}^{\mathcal{T}_{\rm th}^*}\frac{1}{\sqrt{2\pi\delta_2^4/N}}e^{-\frac{\left(q-\delta^2_2\right)^2}{2\delta^4_2/N}}{\rm d} q\notag\\
=&p_1Q\left(\frac{\mathcal{T}_{\rm th}^*-\delta^2_1}{\delta^2_1/\sqrt{N}}\right)+p_2Q\left(\frac{\delta^2_2-\mathcal{T}_{\rm th}^*}{\delta^2_2/\sqrt{N}}\right),
\end{align}
where $Q(x)=\frac{1}{\sqrt{2\pi}} \int^{\infty}_{x}e^{-\frac{1}{2}x^2}{\rm d}x$ is the Q-function.}}
{\section{Capacity Analysis}}
In this section, we first derive the PDF of the received signal and then calculate the information entropies. Subsequently, the channel capacity of the considered jamming channel under the AAJ scheme is derived by maximizing the mutual information over the distribution of the TN transmitted signals.

Based on the definition of the channel capacity in information theory \cite{IEEEhowto:23}, the channel capacity of the jamming channel can be given by
\begin{align}
\mathcal{C}=\max\limits_{p(a_k)} \ {\bf{\rm{I}}}(Y_{{\rm R}}; A)=\max\limits_{p(a_k)} \ {\rm{H}}(Y_{\rm R})-{\rm{H}}(Y_{\rm R}|A),
\end{align}
where ${\bf{\rm{I}}}(Y_{\rm R}; A)$ is the average mutual information between the input $A$ and the output $Y_{\rm R}$. ${\rm{H}}(Y_{\rm R})$ and ${\rm{H}}(Y_{\rm R}|A)$ are information entropies defined as
\begin{align}
{\rm{H}}(Y_{\rm R})&\!=\!-\int_{-\infty}^{\infty} f(y_{_{\rm R}})\log f(y_{_{\rm R}}){\rm d}y_{_{\rm R}},\\
{\rm{H}}(Y_{\rm R}|A)&\!=\!-\!\int_{-\infty}^{\infty}\!\!\int_{-\infty}^{\infty}\! f(a_k)f_{Y_{\rm R}|A}\!(y_{_{\rm R}}|a_k)\!\log f_{Y_{\rm R}|A}(y_{_{\rm R}}|a_k){\rm d}a_k{\rm d}y_{_{\rm R}},
\end{align}
respectively.

To derive the channel capacity of the jamming channel with the AAJ scheme, we first derive the PDF of $Y_{\rm R}$ and $Y_{\rm R}|A$, i.e., $f(y_{_{\rm R}})$ and $f_{Y_{\rm R}|A}(y_{_{\rm R}}|a_k)$, respectively. From (18), the PDF of $Y_{\rm R}|A$ under the Gaussian channel can be given {by}
\begin{align}
f_{{Y_{\rm R}}|A}(y_{_{\rm R}}|a_k)&=\frac{1}{\sqrt{2\pi\delta^2_k}}e^{-\frac{y^2_{{\rm R}}}{2\delta^2_k}}.
\end{align}

Next, we compute the PDF of $Y_{\rm R}$. According to the relationship between the marginal PDF and joint PDF, $f(y_{_{\rm R}})$ can be computed as
\begin{align}
f(y_{_{\rm R}})&=\sum_{k=1}^2 f(y_{_{\rm R}},a_k)\notag\\
&=\sum_{k=1}^2 f_{Y_{\rm R}|A}(y_{_{\rm R}}|a_k)p(a_k).
\end{align}
\begin{Proposition}
Considering the binary input at the TN, i.e., $p_1=p$ and $p_2=1-p$,  the closed-form of the mutual information for the jamming channel with the AAJ scheme is given by
\begin{align}
{\rm I}(Y_{\rm R}; A)=&-\int_{-\infty}^{\infty}  \left(p\varPhi_1 +(1-p)\varPhi_2 \right)\log \left(p\varPhi_1 +(1-p)\varPhi_2 \right){\rm d}y_{_{\rm R}}\notag\\
 &-\frac{1}{2} p\log 2\pi e \delta^{2}_{1}-\frac{1}{2} (1-p)\log 2\pi e \delta^{2}_{2},
\end{align}
where $\varPhi_1=\frac{1}{\sqrt{2\pi\delta^{2}_{1}}}e^{-\frac{y_{_{\rm R}}^2}{2\delta^{2}_{1}}}$ and $\varPhi_2=\frac{1}{\sqrt{2\pi\delta^{2}_{2}}}e^{-\frac{y_{_{\rm R}}^2}{2\delta^{2}_{2}}}$.
\end{Proposition}

\emph{Proof:} Please see Appendix D. \hfill\rule{2mm}{2mm}

Therefore, the channel capacity of the jamming channel can be reformulated as
\begin{align}
 \mathcal{C}=\max\limits_{p} \ {\rm I}(Y_{\rm R}; A).
\end{align}

The  optimal $p$ and the corresponding channel capacity are summarized in the following proposition.
\begin{Proposition}
The channel capacity of the jamming channel is given {by}
\begin{flalign}
\mathcal{C}&=-\int_{-\infty}^{\infty}  \left(p^*\varPhi_1 \!+\!(1-p^*)\varPhi_2 \right)\log \left(p^*\varPhi_1 \!+\!(1-p^*)\varPhi_2 \right){\rm d}y_{_{\rm R}}\notag\\
&\quad\ \!-\frac{1}{2} p^*\log 2\pi e \delta^{2}_{1}-\frac{1}{2} (1-p^*)\log 2\pi e \delta^{2}_{2},
\end{flalign}
where $p^*$ is the optimal input distribution, which is determined by
\begin{align}
\int_{-\infty}^{\infty}(\varPhi_2-\varPhi_1 ) \log (p\varPhi_1+(1-p)\varPhi_2 ) {\rm d}y_{_{\rm R}}=\frac{1}{2} \log\frac{\delta^{2}_{1}}{\delta^{2}_{2}}.
\end{align}
\end{Proposition}

\emph{Proof:} Please see Appendix E.\hfill\rule{2mm}{2mm}

In order to solve the problem (51) to derive the precise value of $p^*$, we denote $\eta(p,y_{_{\rm R}})=(\varPhi_2-\varPhi_1 ) \log (p\varPhi_1+(1-p)\varPhi_2 )$. Therefore, the  first-order derivative of $\int^{\infty}_{-\infty}\eta(p,y_{_{\rm R}}){\rm d}y_{_{\rm R}}$ can be computed as
\begin{align}
    \frac{{\rm d}\int^{\infty}_{-\infty}\eta(p,y_{_{\rm R}}){\rm d}y_{_{\rm R}}}{{\rm d}p}=-\int^{\infty}_{-\infty} \frac{\left(\varPhi_1-\varPhi_2\right)^2}{\ln2\left(p\varPhi_1+(1-p)\varPhi_2 \right)}{\rm d}y_{_{\rm R}}.
\end{align}
Note that $\varPhi_1\neq\varPhi_2$ and $\left(p\varPhi_1+(1-p)\varPhi_2 \right)>0$. Thus, $ {\rm d}\int^{\infty}_{-\infty}\eta(p,y_{_{\rm R}}){\rm d}y_{_{\rm R}}/{\rm d}p< 0$, which means that $\int^{\infty}_{-\infty}\eta(p,y_{_{\rm R}}){\rm d}y_{_{\rm R}}$ monotonically decreases over $p$. Therefore, the solution of  the problem (51), i.e., $\int^{\infty}_{-\infty}\eta(p,y_{_{\rm R}}){\rm d}y_{_{\rm R}}-\frac{1}{2}\log\frac{\delta^2_{\rm 1}}{\delta^2_{\rm 2}}=0$, can be found by bisection search in $[0,1]$.

\section{Numerical Results}
In this section, both the BER and channel capacity performances of the AAJ  scheme are evaluated with numerical simulation results. We define the SNR at the RN as $P_{\rm A}/ \sigma^2_{\rm R}$, where $P_{\rm A}$ is the average power of the desired signal $A(t)$. Unless specifically stated, we assume that the JN launches random broadband jamming signal, i.e., CSCG signal. {Furthermore, it is assumed that there exists a  dominant line-of-sight (LOS) path among each of the following  three links, i.e., JT, TR, and JR links. Therefore, we consider  Rician fading models throughout this section. The channel coefficients of the JT link, the TR link, and the JR link are denoted by  $h_1=\sqrt{\frac{\mathcal{K}}{\mathcal{K}+1}}\overline{h}_1+\sqrt{\frac{1}{\mathcal{K}+1}}\hat{h}_1$, $h_2=\sqrt{\frac{\mathcal{K}}{\mathcal{K}+1}}\overline{h}_2+\sqrt{\frac{1}{\mathcal{K}+1}}\hat{h}_2$, and $h_3=\sqrt{\frac{\mathcal{K}}{\mathcal{K}+1}}\overline{h}_3+\sqrt{\frac{1}{\mathcal{K}+1}}\hat{h}_3$, respectively, where  the Rician factor $\mathcal{K}=10$. $\overline{h}_1$, $\overline{h}_2$, and $\overline{h}_3$ are the LOS components,  which are all set to be $1$.   $\hat{h}_1$, $\hat{h}_2$, and $\hat{h}_3$ are the non-line-of-sight (NLOS) components,  which are generated by the i.i.d. complex Gaussian distribution $\mathcal{CN}(0,1)$.} The length of the preamble sequence is set to be $m=10$. The variance of the CSCG noise is  normalized to $1$, i.e., $\sigma_{\rm R}^2=1$. A total of $10^7$ Monte-Carlo runs are averaged to show the results.

\begin{figure}[htbp]
 \centering
  \includegraphics[width=0.7\textwidth]{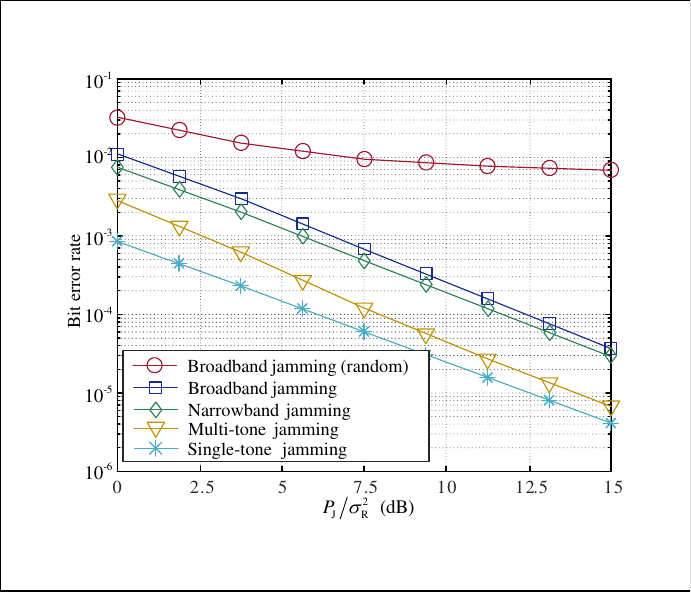}
 \caption{BER comparison among  different unmodulated jamming signals ($P_{\rm A}/\sigma_{\rm R}^2=5$ dB, $N=10$).}
\end{figure}
\begin{figure}[htbp]
 \centering
  \includegraphics[width=0.7\textwidth]{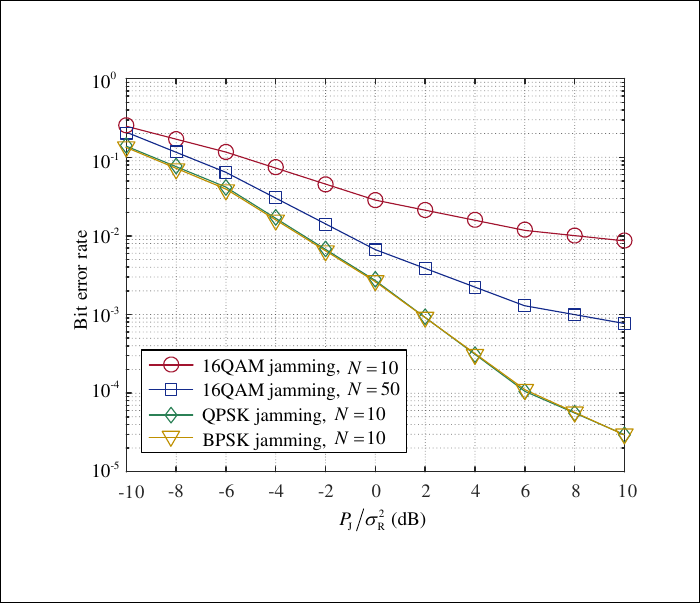}
 \caption{BER comparison among different modulated jamming signals ($P_{\rm A}/\sigma^2_{\rm R}=5$ dB).}
\end{figure}

Figure 2 shows how the BER performance changes with the JNR under different unmodulated jamming signals. Specifically, the broadband jamming signal (random) is generated by a band-limited CSCG signal that  occupies a  bandwidth of $10$ MHz with the center frequency $100$ MHz; the broadband jamming signal (deterministic) is a sum of  sinusoid signals  that also constitutes a  bandwidth of $10$ MHz with the center frequency $100$ MHz; the narrowband jamming is a sum of sinusoid signals that constitutes a  bandwidth of $100$ KHz with the center frequency $100$ MHz; the multi-tone jamming signal is comprised of  five sinusoid signals at frequencies $\{95, 97.5, 100, 102.5, 105\}$ MHz; and the single-tone jamming signal is a sinusoid signal at the frequency of $100$ MHz. The bandwidth of the baseband signal is $10$ MHz. The SNR is set to be $P_{\rm A}/\sigma^2_{\rm R}=5$ dB. The number of samples collected during one symbol is set to be $N=10$. From the figure, we can observe that the AAJ scheme works well even under different jamming attacks. More importantly, the BERs of the considered jamming cases are all monotonically decreasing with the JNR, i.e., the higher power the jammer uses, the better BER performance we can get, which is a great advantage as compared with the existing anti-jamming schemes. As shown in the previous analysis, the  BER of the broadband jamming (random) case is floored  with the increase of the JNR. In addition, the BER of the broadband jamming (random) case is the worst due to the fact that  Gaussian signal is a random signal while the other four cases can be regarded as deterministic ones, which coincides with our analysis in Section IV and  Section V.

{Figure 3 shows the system BER performances under different modulated jamming signals, i.e., BPSK, QPSK, and 16QAM (jamming types). The bandwidth of the jamming signal is $10$ MHz and the carrier frequency is $100$ MHz. The bandwidth of the baseband signal is $1$ MHz. The SNR is set to be $P_{\rm A}/\sigma^2_{\rm R}=5$ dB. It is observed that the BERs under the modulated jamming signals still monotonically decrease with the JNR. In the case of the sampling rate $N=10$, the BER of the 16QAM jamming case performs worst  due to the fact that a higher order modulation incurs higher uncertainty of the jamming power. However, one can see that the BER gets improved when $N$ is increased to $50$.  In addition, the BER of the BPSK  jamming case is close to that of the QPSK one. The reason is that the envelope of  the phase modulated signal is constant, and thus the energy levels of the BPSK and QPSK jamming signals are the same, i.e., the BER performances of these two cases are equivalent under the energy detection criterion.}

\begin{figure}[htbp]
 \centering
  \includegraphics[width=0.72\textwidth]{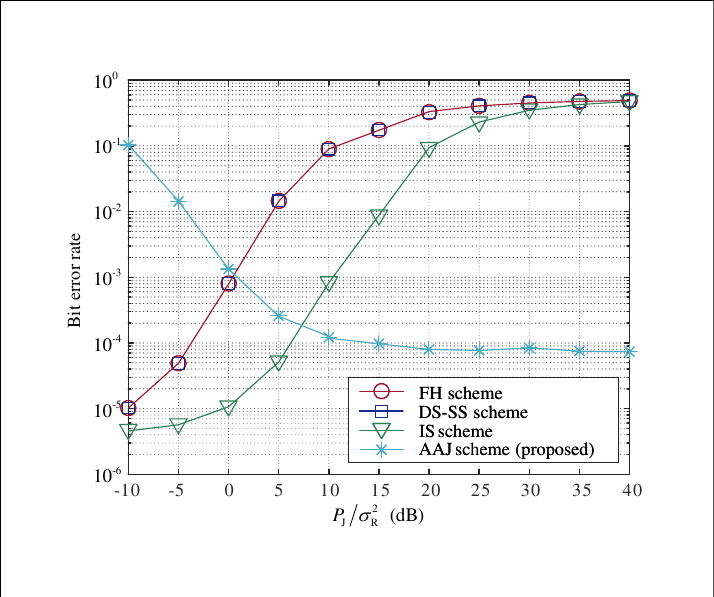}
 \caption{BER comparison among  different anti-jamming schemes ($E_{\rm b}/N_0=10$ dB, $N=8$).}
\end{figure}
\begin{figure}[htbp]
 \centering
  \includegraphics[width=0.7\textwidth]{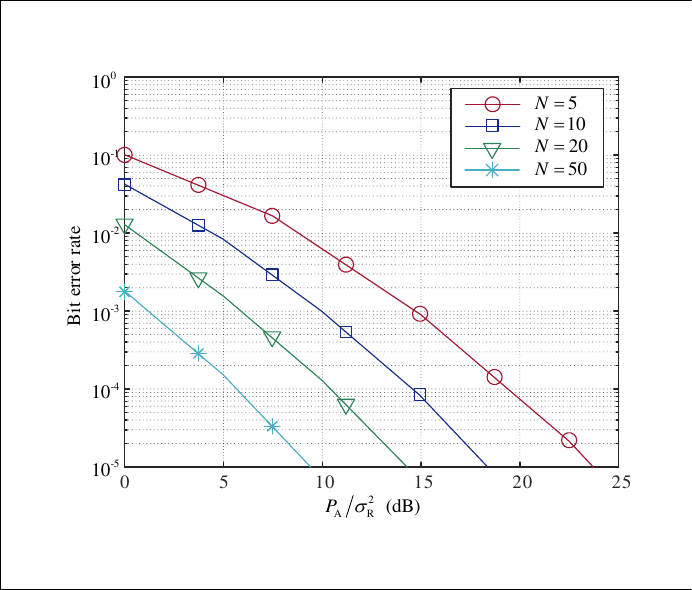}
 \caption{BER comparison with  different sampling rate ($P_{\rm J}/\sigma_{\rm R}^2$=10 dB).}
\end{figure}

Figure 4 illustrates how the BER performance changes with the JNR under different anti-jamming schemes. The broadband (fullband) jamming attack, i.e., CSCG signal with $\mathcal{CN}(0,P_{\rm J})$,  is adopted in this simulation. For a fair comparison, the SNR per bit, also know as $E_{\rm b}/N_{\rm 0}$ (ratio of the energy per bit to the noise power spectral density), is set to be $10$ dB. The spread factor in the DS-SS and FH schemes are both set to be $8$. Likewise, we adopt the number of  samples collected during one symbol as $N=8$. Since the conventional anti-jamming schemes require the instantaneous CSI, the channel coefficients are all set to be $1$, i.e., $h_1=h_2=h_3=1$. From the figure, it can be observed that the BER under our proposed AAJ scheme decreases with the JNR while those of the existing schemes increase with the JNR. This is because the jamming signal is used as a carrier of the useful messages in  the AAJ scheme, i.e., it can enhance the transmission of the TR link. When the JNR $P_{\rm J}/\sigma_{\rm R}^2=40$ dB, the BER of the AAJ scheme is about $7.38\times10^{-5}$. In this case, however, the BERs of the other schemes decrease to $0.5$, thus failing to decode out any meaningful messages.

Figure 5 illustrates the BER performances of the different numbers of samples, $N$. The JNR is set to be $P_{\rm J}/\sigma_{\rm R}^2=10$ dB.  It is observed that the BER of the AAJ scheme significantly decreases with $N$. Therefore, one can not only increase the SNR but also the sampling rate to improve the transmission reliability of the legitimate communication. In this system, since the jamming signal is unknown, it is desirable to increase the interval of one symbol due to the fact that the longer the interval of one symbol is, the average energy of one symbol approaches to a constant. In other words, the bandwidth of the legitimate transmission should be narrower than the jamming signal even though  it may decrease the transmission rate. Therefore, there is a tradeoff between the transmission rate and the BER performance in the AAJ scheme.
\begin{figure}[htbp]
 \centering
  \ \includegraphics[width=0.7\textwidth]{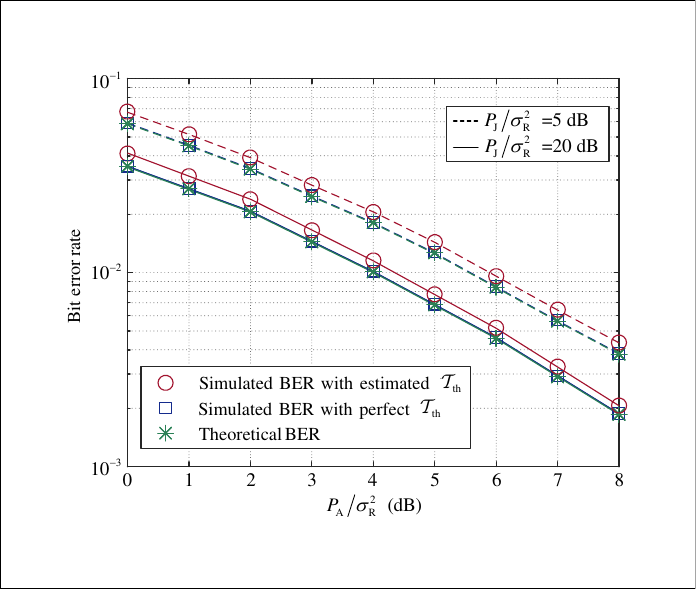}
 \caption{BER comparison with the simulated and theoretical results ($m=10$, $N=10$).}
\end{figure}

\begin{figure}[htbp!]
 \centering
 \includegraphics[width=0.705\textwidth]{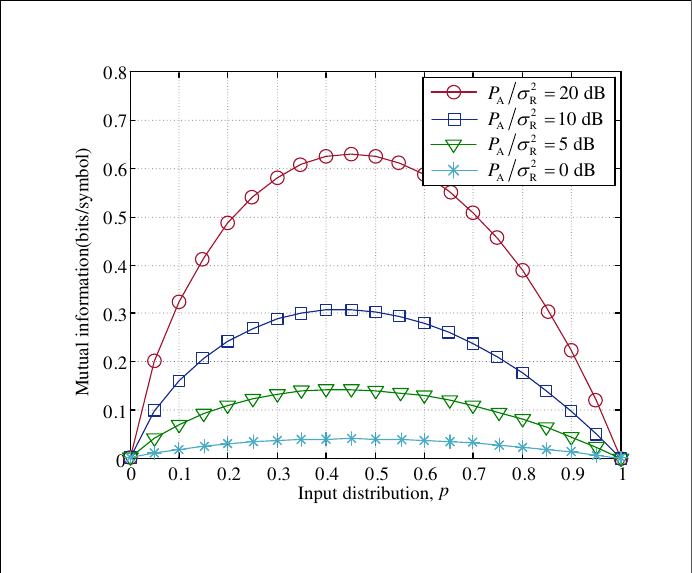}
 \caption{Mutual information \emph{v.s.} input distribution in the jamming channel under the AAJ scheme.}
\end{figure}
  \begin{figure}[htbp!]
 \centering
  \includegraphics[width=0.705\textwidth]{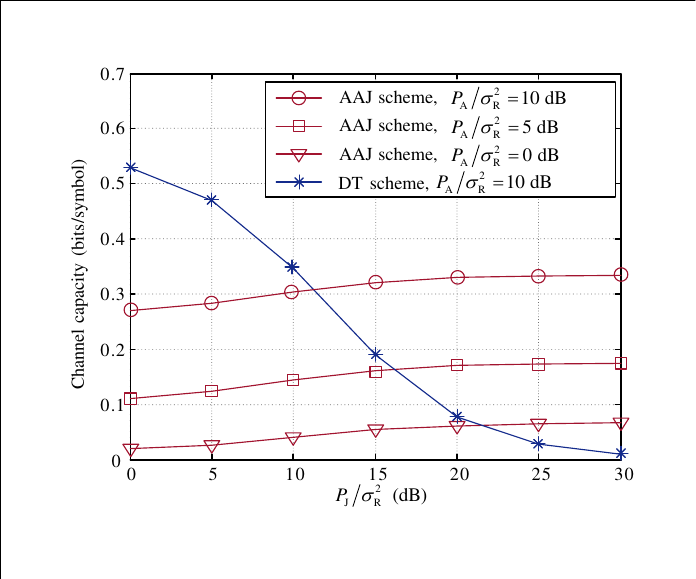}
 \caption{Channel capacity \emph{v.s.} JNR.}
\end{figure}

The simulated and theoretical BERs are compared in Fig. 6.  In this figure, $N=10$.  For the case of the simulated BER with the estimated  $\mathcal{T}_{\rm th}$, the length of the preamble sequence $m=10$; For  the case of the  simulated BER with the perfect $\mathcal{T}_{\rm th}$, the  detection threshold  $\mathcal{T}_{\rm th}$ is perfect estimated, i.e., $\mathcal{T}^*_{\rm th}$ is  known. It is observed that the simulated BER coincides with the theoretical result if   $\mathcal{T}^*_{\rm th}$ is known. Even when we use short preamble sequences, the BER is still close to the case where $\mathcal{T}_{\rm th}$ is perfectly known, i.e., the system can still achieve a good BER performance.

The relationship between the mutual information and the input distribution for the Gaussian jamming channel is shown in Fig. 7. {In this simulation, the JNR is set to be  $P_{\rm J}/\sigma^2_{\rm R}=10$ dB, whereas the SNRs are drawn from $\{0,5,10,20\}$ dB.} It is obvious that the mutual information between $Y_{\rm R}$ and $A$ is a concave function over the input distribution, and there is only one maximum value for $p\in[0,1]$, thus verifying our previous analysis. It is also interesting to observe that the optimal input distribution approaches a uniform distribution, i.e., $p=0.5$, when the SNR increases. In addition, it is shown that the mutual information increases with the SNR, and thus we can still increase the SNR to improve the transmission performance.

Figure 8 shows the relationship between the channel capacity and the JNR of the AAJ scheme. From the figure, it can be observed that the channel capacity of the DT scheme decreases with the JNR while the ones for the AAJ scheme increase with the JNR. For  the SNR $P_{\rm A}/\sigma^2_{\rm R}=10$ dB, when the JNR is relatively large, i.e., $P_{\rm J}/\sigma_{\rm R}^2>11.8$ dB, the AAJ scheme begins to exceed the DT one, meaning that the proposed AAJ scheme has a great advantage in the strong jamming signal case. In addition, the channel capacity monotonically increases with the SNR.
\section{Conclusions and Future Works}
In this paper, we have proposed  a novel AAJ  scheme for a jammed channel to maintain the legitimate communications between the TN and RN. By exploiting the JM method, the TN is able to re-modulate the jamming signals with its own messages for new anti-jamming communications.  Focusing on the binary input at the TN, we have computed the  minimum average BER of the AAJ scheme and the corresponding optimal detection threshold, where the single-tone, multi-tone, narrowband, and broadband jamming signals were all considered. Then, the semi-closed forms of both the channel capacity and the corresponding optimal input distribution of the jammed channel with the AAJ scheme  have been derived. Simulation results have been presented to show that the proposed anti-jamming scheme is effective even under the extremely strong and/or broadband jamming attacks. When the JNR is relatively high, the AAJ scheme outperforms the DT one in terms of the channel capacity. Furthermore, we have revealed that there exists a BER floor with increasing JNR  through the asymptotic analysis. Finally, it is interesting to point out that the RN enjoys increasingly improved error rate performance for larger JNR, making it an ideal anti-jamming candidate when the jamming attacks are hard to be suppressed.

As a pioneering work, we believe that the proposed  AAJ scheme provides a new solution for anti-jamming communications. In the future, we plan to expand our work from the following perspectives: 1) Mode selection: when the jamming signal is weak or absent, the AAJ scheme may need more power to guarantee the quality of the legitimate communications. It is worthy to develop a hybrid transmission scheme, e.g., how to select the AAJ, DT, or conventional scheme according to the different jamming cases; 2) Trade-off between the BER and the transmission rate: as mentioned in Section VII, a longer symbol interval brings a lower BER but leads to reduced  transmission rate at the same time. It is therefore of great interest to investigate the relationship between BER and transmission rate; 3) Multi-antenna case: in the present research, the TN and RN are equipped with one single antenna. Since multiple antennas can provide a  higher diversity gain, further research is needed to study the performance limits of the AAJ scheme in the multi-antenna case; { 4) Optimal detection method: optimal/enhanced detection schemes for the proposed  AAJ scheme are a vital and interesting topic. For example,  maximum likelihood (ML) detection and independent component analysis (ICA), are worthy of a close investigation; 5)  Power-varying jamming cases: in practice, the jammer does not need to always send signals with fixed power, and thus the performance analysis of the AAJ scheme under the power-varying jamming cases will be studied in our future work.}
{\section*{Acknowledgement}
The authors are deeply indebted to the Editor Dr. Giovanni Geraci and  the anonymous Reviewers for many of their insightful comments which have greatly helped improve the quality of this work.}
\appendices

\section{Proof of \emph{Lemma 3.1}}
According to  probability theory and statistics \cite{IEEEhowto:24}, the chi-square distribution with $l$ degrees  of freedom is defined as the distribution of  a sum of squares of $l$ independent standard normal random variables. {Note that $y_{_{\rm R}}[n]$ is a CSCG variable with zero mean and variance $\delta^2_k$, i.e., $y_{_{\rm R}}[n]=r+jr$, where $r\sim\mathcal{N}\left(0,\delta^2_k/2\right)$. Therefore, the average energy of one symbol, $\mathcal{Q}= \frac{1}{N} \sum^N_{n=1} |y_{_{\rm R}}[n]|^2$ $=\sum_{n=1}^{2N}(r/\sqrt{N})^2$, can be regarded as a sum of squares of $2N$ CSCG variables with distribution $r/\sqrt{N}\sim$ $\mathcal{CN}(0,\delta^2_k/2N)$.

By normalizing the variance of $r/\sqrt{N}$, we have
\begin{align}
\mathcal{Q}&=\frac{1}{N} \sum^N_{n=1} |y_{_{\rm R}}[n]|^2\notag\\
\longrightarrow\frac{2N}{\delta^2_k}\mathcal{Q}&=\sum^N_{n=1} \left|\frac{\sqrt{2}y_{_{\rm R}}[n]}{\delta_k}\right|^2,
\end{align}
where $\sqrt{2}r/\delta_k$  satisfies the standard normal distribution, i.e.,}
\begin{align}
\frac{\sqrt{2}}{\delta_k}r\sim\mathcal{N}(0,1).
\end{align}

Then, we can further derive the following result
\begin{align}
\left|\frac{\sqrt{2}y_{_{\rm R}}[n]}{\delta_k}\right|^2&\sim \mathcal{X}^2(2)\notag\\
\longrightarrow \sum^N_{n=1} \left|\frac{\sqrt{2}y_{_{\rm R}}[n]}{\delta_k}\right|^2&\sim\mathcal{X}^2(2N).
\end{align}

Therefore, $(2N\mathcal{Q})/\delta^2_k$ is a chi-square distributed variable with $2N$ degrees of freedom. Hence, \emph{Lemma 3.1} is proved.

\section{The Proof of \emph{Lemma 3.2}}
From \emph{Lemma 3.1}, it is found out that $X=(2N\mathcal{Q})/\delta^2_k$ is a chi-square distributed variable with $2N$ degrees of freedom, whose PDF is given {by}
\begin{align}
f_X(x)=\left\{
             \begin{array}{lr}
              \frac{1}{2^N\Gamma(N)} \left(x\right)^{N-1} e^{-\frac{x}{2}},  & x>0,\\
           0, & \text{otherwise}.
             \end{array}
             \right.
\end{align}

Next, we compute the PDF of $\mathcal{Q}$ according to the property of a function of random variables. Denote $F_X(x)$ and $F_{\mathcal{Q}(a_k)}(q)$ as the cumulative density functions (CDF) of $X$ and $\mathcal{Q}$.

Therefore, we have  $\mathcal{Q}=\left(\delta^2_k X\right)/2N$, and then $F_{\mathcal{Q}}(q)$ is given {by}
\begin{align}
F_{\mathcal{Q}}(q)&={\rm Pr}\{\mathcal{Q}\leq q\}\notag\\
&={\rm Pr}\left\{X\leq\frac{2Nq}{\delta^2_k}\right\}\notag\\
&=F_X\left(\frac{2Nq}{\delta^2_k}\right).
\end{align}

By computing the first-order derivative of the CDF (59), we can derive its PDF as follows
\begin{align}
f_\mathcal{Q}(q)=\frac{{\rm d}F_{\mathcal{Q}}(q)}{{\rm d}q}&=\frac{2N}{\delta^2_k}f_X(\frac{2Nq}{\delta^2_k}).
\end{align}
Therefore, by substituting (58) into (60), we have
\begin{align}
f_\mathcal{Q}(q)=\left\{
             \begin{array}{lr}
              \frac{N}{\Gamma(N)\left(\delta^2_k\right)^N} \left(N q\right)^{N-1} e^{-\frac{N}{\delta^2_k}q},  & q>0,\\
           0, & \text{otherwise}.
             \end{array}
             \right.
\end{align}

Hence, \emph{Lemma 3.2} is proved.
\section{The Proof of \emph{Proposition 3.1}}

To compute the minimum average  BER of the energy detection under the AAJ scheme, we first need to determine the optimal detection threshold.

First, by substituting the PDF $f_\mathcal{Q}(q)$ into (16), the BER can be rewritten as
\begin{align}
P_e=&\sum^2_{k=1} p_k (1-\int_{\mathcal{I}_k} f_\mathcal{Q}(q)){\rm d}q\notag\\
=&p_1 \left(1-\int^{\mathcal{T}_{\rm th}^*}_0  \frac{N}{\Gamma(N)\left(\delta^2_1\right)^N} \left(N q\right)^{N-1} e^{-\frac{N}{\delta^2_1}q} \right){\rm d}q\notag\\
&+p_2 \left(1-\int_{\mathcal{T}_{\rm th}^*}^{+\infty}  \frac{N}{\Gamma(N)\left(\delta^2_2\right)^N} \left(N q\right)^{N-1} e^{-\frac{N}{\delta^2_2}q} \right){\rm d}q.
\end{align}
Then, we compute the first-order derivative of $P_e$, i.e.,
\begin{align}
\frac{{\rm d}P_e}{{\rm d}\mathcal{T}_{\rm th}}=&p_1\times \frac{N}{\Gamma(N)\left(\delta^2_1\right)^N} \left(N \mathcal{T}_{\rm th}\right)^{N-1}e^{-\frac{N}{\delta^2_1}\mathcal{T}_{\rm th}}\notag\\
&-p_2\times\frac{N}{\Gamma(N)\left(\delta^2_2\right)^N} \left(N \mathcal{T}_{\rm th}\right)^{N-1} e^{-\frac{N}{\delta^2_2}\mathcal{T}_{\rm th}}.
\end{align}

When ${\rm d}P_e/{\rm d}\mathcal{T}_{\rm th}=0$, by performing some algebraic manipulation, we can derive the expression of the optimal detection threshold $\mathcal{T}^*_{\rm th}$, {as} presented in (24).

Therefore, the minimum average BER of  the energy detection under the AAJ scheme is derived and given in (23).

Hence, \emph{Proposition 3.1} is proved.

\section{Proof of \emph{Proposition 6.1}}
 According to (45), to derive the mutual information for the jamming channel, we need to compute the information entropies ${\rm{H}}(Y_{\rm R})$ and ${\rm{H}}(Y_{\rm R}|A)$.

From (47) and (48), the information entropy of $Y_{\rm R}|A=a_k$ can be computed as
\begin{align}
{\rm H}(Y_{\rm R}|A=a_k)=&-\int_{-\infty}^{\infty} f_{Y_{\rm R}|A}(y_{_{\rm R}}|a_k)\log f_{Y_{\rm R}|A}(y_{_{\rm R}}|a_k){\rm d}y_{_{\rm R}}&\notag\\
=&-\int_{-\infty}^{\infty} \frac{1}{\sqrt{2\pi\delta^{2}_{k}}}e^{-\frac{y^2_{\rm R}}{2\delta^{2}_{k}}} \log\left(\frac{1}{\sqrt{2\pi\delta^{2}_{k}}}e^{-\frac{y^2_{\rm R}}{2\delta^{2}_{k}}}\right){\rm d}y_{_{\rm R}}\notag\\
=&- \int_{-\infty}^{\infty} \frac{1}{\sqrt{2\pi\delta^{2}_{k}}}e^{-\frac{y^2_{\rm R}}{2\delta^{2}_{k}}}\left(\log\frac{1}{\sqrt{2\pi\delta^{2}_{k}}}-\frac{y^2_{\rm R}\log e}{2\delta^{2}_{k}}\right){\rm d}y_{_{\rm R}}\notag\\
=&\frac{1}{2}\log 2\pi\delta^{2}_{k}+\frac{\mathbb{E}\left[Y^2_{\rm R}\right]\log e}{2\delta^{2}_{k}}\notag\\
=&\frac{1}{2}\log 2\pi e \delta^{2}_{k}.
\end{align}

Further, the information entropy of  $Y_{\rm R}|A$ is given {by}
\begin{align}
{\rm H}(Y_{\rm R}|A)&=\sum_{k=1}^2 p(a_k){\rm H}(Y_{\rm R}|A=a_k)\notag\\
&=\frac{1}{2}\sum_{k=1}^2 p(a_k)\log 2\pi e \delta^{2}_{k}.
\end{align}

Likewise, from (46), (48), and (49), the information entropy of  $Y_{\rm R}$ is given {by}
\begin{flalign}
{\rm{H}}(Y_{\rm R})&=-\int_{-\infty}^{\infty} f(y_{_{\rm R}})\log f(y_{_{\rm R}}){\rm d}y_{_{\rm R}}\notag\\
&=-\int_{-\infty}^{\infty} \sum_{k=1}^2 f_{Y_{\rm R}|A}(y_{_{\rm R}}|a_k)p(a_k)\log \left(\sum_{k=1}^2 f_{Y_{\rm R}|A}(y_{_{\rm R}}|a_k)p(a_k) \right){\rm d}y_{_{\rm R}}\notag\\
&=-\int_{-\infty}^{\infty}\sum_{k=1}^2 p(a_k)\frac{1}{\sqrt{2\pi\delta^{2}_{k}}}e^{-\frac{y^2_{{\rm R}}}{2\delta^{2}_{k}}} \log \left(\sum_{k=1}^2 p(a_k)\frac{1}{\sqrt{2\pi\delta^{2}_{k}}}e^{-\frac{y^2_{{\rm R}}}{2\delta^{2}_{k}}}\right){\rm d}y_{_{\rm R}}.
\end{flalign}

Substituting (65) and (66) into (45), and thus the mutual information between $Y_{\rm R}$ and $A$ for the considered jamming channel with the AAJ scheme is given in (52).

Therefore, this proposition is proved.
\section{Proof of \emph{Proposition 6.2}}
To derive the channel capacity, i.e., the maximum mutual information ${\rm I}(Y_{{\rm R}}; A)$, we first compute the first-order derivative of  ${\rm I}(Y_{{\rm R}}; A)$ as follows
\begin{align}
\frac{{\rm d} {\rm I}(Y_{{\rm R}}; A)}{{\rm d}p}=&-\int_{-\infty}^{\infty}  \left(\varPhi_1 -\varPhi_2 \right)\log \left(p\varPhi_1 +(1-p)\varPhi_2 \right)+\frac{1}{\ln2}(\varPhi_1 -\varPhi_2 ){\rm d}y_{_{\rm R}}\notag\\
&-\frac{1}{2} \log 2\pi e\delta^{2}_{1}+\frac{1}{2}\log 2\pi e \delta^{2}_{2}\notag\\
=&-\int_{-\infty}^{\infty}(\varPhi_1-\varPhi_2) \log (p\varPhi_1+(1-p)\varPhi_2 ){\rm d}y_{_{\rm R}}-\frac{1}{2} \log\frac{\delta^{2}_{1}}{\delta^{2}_{2}}.
\end{align}

Since  ${\rm I}(Y_{{\rm R}}; A)$ is continuous and bounded for  $p\in[0,1]$,  there must exist a maximum value in the feasible set. Moreover, ${\rm I}(Y_{{\rm R}}; A)|_{p=0}={\rm I}(Y_{{\rm R}}; A)|_{p=1}=0$ and $\frac{{\rm d}{\rm I}(Y_{{\rm R}}; A)}{{\rm d}p}\vert_{p=0}>0$,  and thus   the maximum ${\rm I}(Y_{{\rm R}}; A)$  is achieved at the stationary point of the objective function in (51), which is equivalent to the root of equation (67), i.e., ${\rm d}{\rm I}(Y_{{\rm R}}; A)/{\rm d}p=0$.

Therefore, the mutual information of the jamming channel achieves the maximum value when $p_1=p^*$ and $p_2=1-p^*$.

Hence, \emph{Proposition 6.2} is proved.

\end{document}